\documentclass[a4paper]{article}%
\usepackage{blindtext}
\usepackage{geometry}
\usepackage{hyperref}
\hypersetup{
colorlinks=true,
urlcolor=blue,
citecolor=blue}
\usepackage[all]{xy,xypic}
\usepackage{amsfonts,amssymb,amsmath,amsgen,amsopn,amsbsy,theorem,graphicx,epsfig}
\usepackage{eufrak,amscd,bezier,latexsym,mathrsfs,enumerate,multirow}
\usepackage[utf8]{inputenc}
\usepackage[english]{babel}
\usepackage[dvipsnames]{xcolor}
\usepackage{academicons}
\definecolor{orcidlogocol}{HTML}{A6CE39}
\usepackage[pagewise]{lineno}
\usepackage{epstopdf}
\usepackage{xspace}
\usepackage{color}
\usepackage{units}
\usepackage{slashed}
\usepackage{gensymb}
\usepackage{booktabs}
\usepackage{xcolor}
\usepackage{setspace}
\usepackage{physics}
\usepackage{nicefrac}
\usepackage{bbold}
\usepackage{accents}
\usepackage{mathtools}
\usepackage{multirow}

\def\bt{\begin{equation}}
\def\bea{\begin{eqnarray}}
\def\ee{\end{equation}}
\def\eea{\end{eqnarray}}

\begin{document}

\title{Higgs production at next generation $e^{+}e^{-}$ colliders} 
 
\author{ \textbf{Deniz YILMAZ$^{1}$\thanks{dyilmaz@eng.ankara.edu.tr} , Mehmet SAHIN$^{2}$, Dogukan Hazar YAVUZ$^{1}$}\\ 
$^{1}$Physics Engineering Department, Ankara University, Ankara,Turkey \\ 
$^{2}$Department of Computer Engineering, Usak University, Usak,Turkey}

\maketitle

\begin{abstract}
In this study, Higgs production processes, Higgsstrahlung and vector boson (W and Z) fusion processes, were investigated for four different future lepton colliders (CEPC, ILC, CLIC, and FCC-ee). The cross sections for each production process and corresponding backgrounds were calculated considering the ISR and beamstrahlung effects. Various cuts and the b-tagging method were used to reduce the background.  Finally, the number of events for each collider was determined, and significance calculations were performed. In our calculations, high event numbers were obtained for all four colliders for the Higgsstrahlung, W, and Z fusion process. This shows that electron-positron colliders will play an important role in future Higgs physics research.

\end{abstract}

\section{Introduction}
\label{section1}
The discovery of the Higgs boson at the Large Hadron Collider (LHC)  \cite{citation1, citation2} confirmed the electroweak symmetry breaking mechanism of the Standard Model (SM) \cite{citation3,citation4,citation5,citation6}. However, there is still some unknown about the observed Higgs boson: is it the fundamental scalar of the SM, or a more complex object, or part of an extended Higgs sector? Studying the properties of the Higgs boson at the LHC and in future colliders is crucial to understanding its true nature. Up to now, some properties of the Higgs boson have been measured at the LHC with an accuracy of about 10\%  \cite{citation7,citation8,citation9,citation10}. Although the LHC Run 2 to be developed will examine it with higher data, because of the complexity of the internal structure of the proton, the LHC will not be sensitive enough to examine the properties of Higgs. 

Electron-positron colliders, which will be installed to precisely measure the properties of the Higgs particle, have unique capabilities for the measurement of Higgs boson parameters, including the Higgs total cross section, decay width, branching ratios, Higgs width, and determination of Higgs couplings.

Therefore, today, four $e^{+}e^{-}$ colliders are being designed to study the properties of the Higgs boson and other standard model (SM) particles with high precision: the International Linear Collider (ILC) \cite{citation11}, with a center of mass energy of 250 – 500 GeV, Compact Linear Collider (CLIC) \cite{citation12} with center of mass energies of 380 – 1500 – 3000 GeV, Circular Electron Positron Collider (CEPC) with center of mass energies between 90 and 250 GeV \cite{citation13} and the Future $e^{+}e^{-}$ Circular Collider (FCC-ee) \cite{citation14}, which will be located in a new tunnel at CERN at 240 GeV center of mass energy. The main beam parameters of these colliders \cite{citation11,citation12,citation13,citation14} are given in Table~\ref{table1}. The integrated luminosities given here are annual values.  
\begin{table}[h]
\caption{The main collider parameters}
\medskip
\centering\renewcommand{\arraystretch}{1.2}
\begin{tabular}{|l|l|l|ll|lll|}
\hline
\multicolumn{1}{|c|}{Parameters} & \multicolumn{1}{c|}{CEPC} & \multicolumn{1}{c|}{FCC-ee} & \multicolumn{2}{c|}{ILC}              & \multicolumn{3}{c|}{CLIC}                                                         \\ \hline
Center of mass energy (GeV)           & 240                       & 240                         & \multicolumn{1}{l|}{250}    & 500     & \multicolumn{1}{l|}{380}          & \multicolumn{1}{l|}{1500}        & 3000       \\ \hline
Number of particles per bunch  ($10^{10}$)  & 15                        & 18                          & \multicolumn{1}{l|}{2}      & 2       & \multicolumn{1}{l|}{0.52}         & \multicolumn{1}{l|}{0.37}        & 0.37       \\ \hline
Horizontal beam size at IP ($\sigma_x$) ($\mu$m)  & 20.9                 & 13.7                  & \multicolumn{1}{l|}{0.516} & 0.474 & \multicolumn{1}{l|}{0.149} & \multicolumn{1}{l|}{0.06} & 0.04 \\ \hline
Vertical beam size at IP ($\sigma_y$) (nm)   & 60                & 36                  & \multicolumn{1}{l|}{7.66} & 5.86 & \multicolumn{1}{l|}{2.9} & \multicolumn{1}{l|}{1.5} & 1 \\ \hline
Bunch length (mm)                    & 4.4                       & 5.3                         & \multicolumn{1}{l|}{0.3}    & 0.3     & \multicolumn{1}{l|}{0.07}         & \multicolumn{1}{l|}{0.044}       & 0.044      \\ \hline
Luminosity ($10^5 pb^{-1}$ )                     & 6                         & 17                           & \multicolumn{1}{l|}{1.35}      & 1.8       & \multicolumn{1}{l|}{1.5}            & \multicolumn{1}{l|}{3.7}           & 5.9          \\ \hline
\end{tabular}
\label{table1}
\end{table}

In the electron-positron collider, Higgs bosons are produced by the Higgsstrahlung and vector boson (W and Z) fusion processes \cite{citation15, citation16,citation17, citation18,citation19, citation20,citation21}. In this study, these three processes were examined and calculations were performed using CalcHEP \cite{citation22, citation23}. In the electron-positron collider, it is important to consider the effects of ISR and Beamstrahlung  \cite{citation24, citation25}. The parameters listed in Table~\ref{table1} were used to calculate the ISR and the beamstraghlung effects. In section 2 cross sections are given for these three processes. Section 3 provides signal and background analyses, the number of events for each collider, and the significance calculations. Finally, conclusion is provided in the section 4.  

\section{Higgs Production at the electron - positron colliders}
\label{section2} 
The main production processes of Higgs at the $e^{+}e^{-}$ colliders are the Higgsstrahlung and W/Z fusion mechanisms given below, as shown in Figure \ref{figure1}. 
\begin{align*}
&Higgs-strahlung && e^{+}e^{-} \rightarrow ZH\\
&W fusion && e^{+}e^{-} \rightarrow \overline{\nu_{e}} \nu_{e} H \\
&Z fusion && e^{+}e^{-} \rightarrow e^{+}e^{-} H
\end{align*}
\begin{figure}[t!]
\centering
\includegraphics[width=14cm]{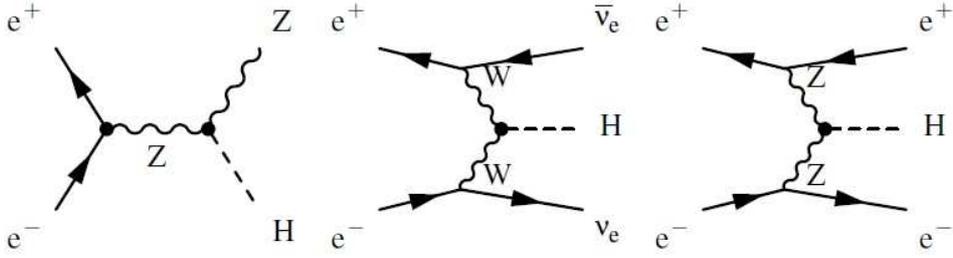}
\caption{The Feynmann diagrams of the Higgs production processes}
\label{figure1}
\end{figure}
\begin{figure}[!b] 
\centering
\includegraphics[width=0.48\textwidth]{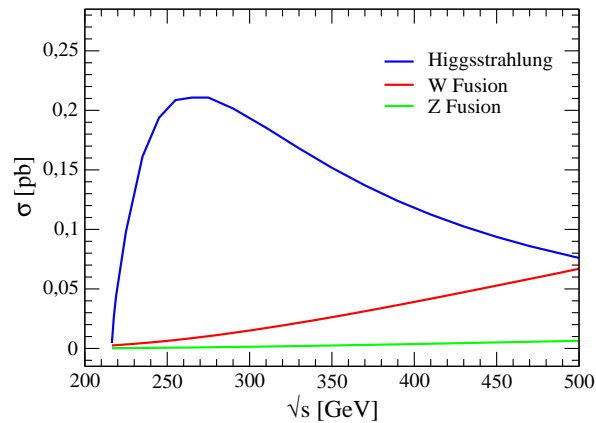}
\caption{The cross sections of
the Higgs production mechanisms as a function of center-of-mass energy.}
\label{figure2}
\end{figure}
The cross section for the Higgsstrahlung process can be written as  
\begin{equation}
\label{equation1} 
\sigma(e^{+}e^{-}\rightarrow ZH )=\dfrac{G^2_F M^4_Z}{96\pi s}(\eta^2_e +a^2_e)\kappa^{1/2}\dfrac{\kappa+12M^2_Z/s}{(1-M^2_Z /s)^2}
\end{equation}
where $a_e = -1$, $\eta_e = -1 +4sin^2 \theta_W$ are the Z charges of the electron and $\kappa = (1-(M_H + M_Z)^2/s)(1-(M_H-M_Z)^2/s)$ is the usual two particle phase space function. 
The total cross section for the vector boson fusion mechanism is  
\begin{equation}
\label{equation1} 
\sigma(e^{+}e^{-}\rightarrow VV \rightarrow l\overline{l}H )=\dfrac{G^2_F m^4_V}{64\sqrt{2}\pi^3 }\int_{x_H}^{1} dx \int_{x}^{1} \dfrac{dyT(x,y)}{[1+(y-x)/x_V]^2}.
\end{equation}
\begin{equation*}
T(x,y)=(\dfrac{2x}{y^3}-\dfrac{3x+1}{y^2}+\dfrac{x+2}{y}-1)[\dfrac{z}{z+1}-log(z+1)]+\dfrac{xz^2(1-y)}{y^3(z+1)},
\end{equation*}
where $V$ denotes the vector bosons W or Z and $x_H=m^2_H/s$, $x_V=m^2_V/s$ and $z=y(x-x_H)/xx_V$ ($\sqrt{s}$ is the center-of-mass energy) \cite{citation26}.

\begin{figure}[!t]
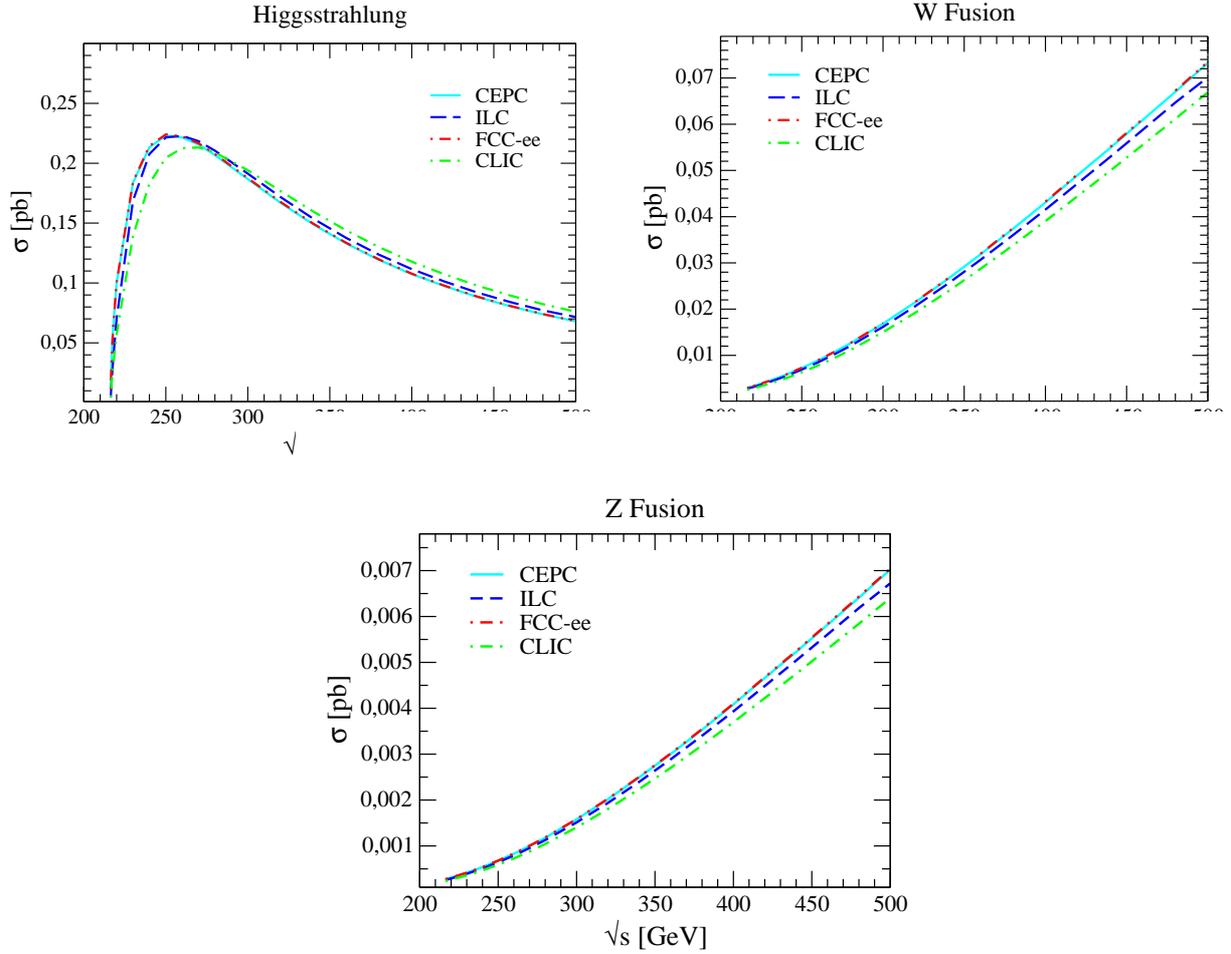

\centering
\begin{minipage}{0.48\textwidth}
\includegraphics[width=\textwidth]{figure3_1.eps}
\end{minipage}\hfill
\begin{minipage}{0.48\textwidth}
\includegraphics[width=\textwidth]{figure3_2.eps}
\end{minipage}\par
\vskip\floatsep
\includegraphics[width=0.48\textwidth]{figure3_3.eps}
\caption{The cross section comparison for Higgsstrahlung, W Fusion and Z Fusion processes for four $e^+e^-$colliders.}
\label{figure3}
\end{figure}

The behavior of the production cross-sections of the Higgs boson calculated by the Higgsstrahlung and the W/Z fusion mechanisms using the CalcHEP simulation program, depending on the center of mass energy, are shown in Figure \ref{figure2} and  Figure \ref{figure3}. The relevant production cross sections as a function of the center of mass energy are shown in Figure \ref{figure2}. As shown in Figure \ref{figure2}, the Higgsstrahlung suppresses the  vector boson production processes for moderate values of the energy due to the additional electroweak coupling. With the increase in energy, the cross sections of the vector boson procecesses increase logarithmically and become dominant.  At a center of mass energy of about 250 GeV, Higgs bosons are predominantly produced from the ZH process as seen in the same figure. In the Figure \ref{figure3}, the cross sections are shown as a function of the center of mass energy for each production mechanisms for four electron-positron colliders with the ISR and the beamstrahlung effects of each colliders.
\section{Signal and Background Analyses}
\label{section3} 

\begin{figure}[!b]
\centering
\begin{minipage}{0.48\textwidth}
\includegraphics[width=\textwidth]{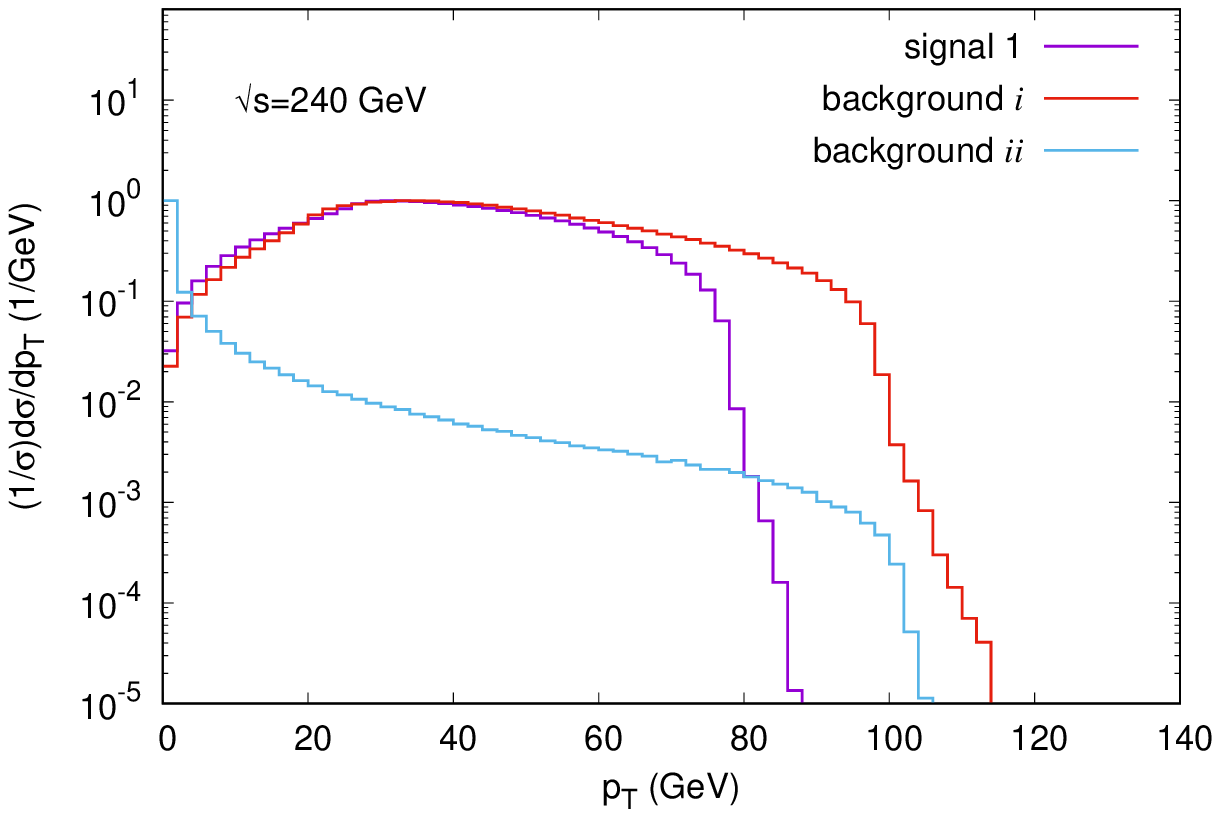}
\end{minipage}\hfill
\begin{minipage}{0.48\textwidth}
\includegraphics[width=\textwidth]{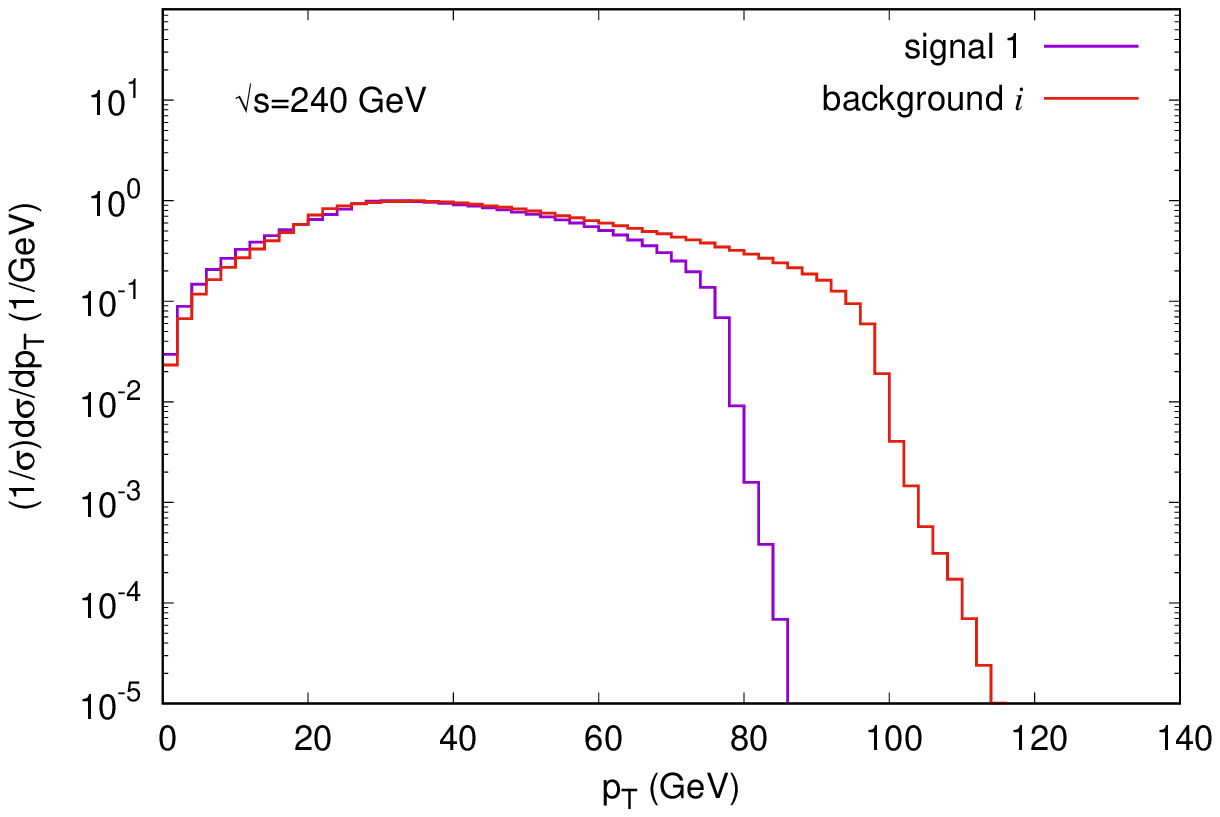}
\end{minipage}\par
\vskip\floatsep
\includegraphics[width=0.48\textwidth]{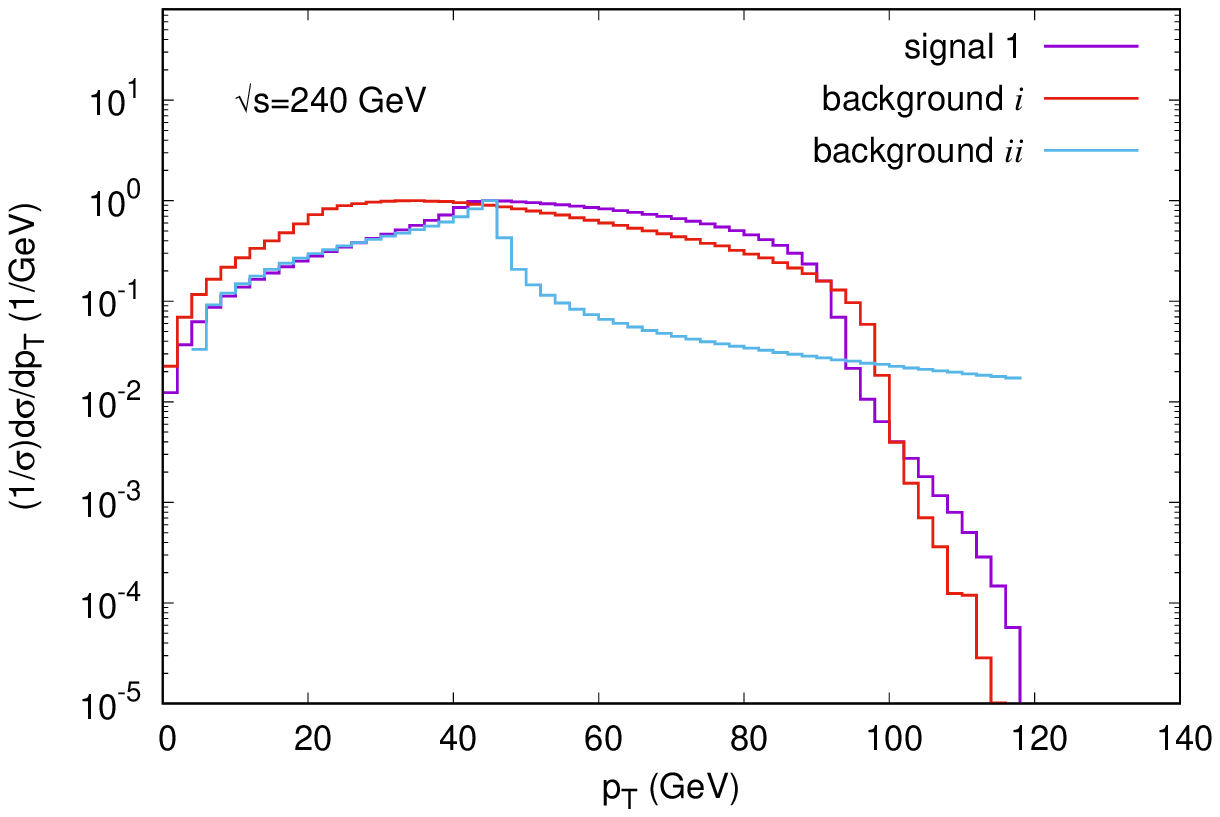}
\caption{Transverse momentum distribution plots for the $e^-/e^+$ (upper left ), $\mu^-/\mu^+$ (upper right) and $J/J$ (bottom) final state particles of signal 1 and the corresponding bacground processes in FCC-ee collider with 240 GeV center of mass energy. }
\label{figure4}
\end{figure}

Because the Higgs boson's decay rate to $b\overline{b}$ is greater than the decay rate to other quarks and leptons \cite{citation27,citation28,citation29,citation30,citation31}, the $b\overline{b}$ decay mode of Higgs ($H \longrightarrow b\overline{b}$) is considered in all production processes in this study. Since the cross sections of the background processes corresponding to the leptonic decays of the Z boson are less than the background cross sections corresponding to the other decays, the leptonic decays of the Z boson in the Higgsstrahlung process are taken into account. The signal processes are given below. 
\begin{align*}
& \text{Signal 1:} \\
& \quad \quad Higgsstrahlung && e^{+}e^{-} \rightarrow ZH \rightarrow l \overline{l} b\overline{b} \\
& \quad \quad Z fusion && e^{+}e^{-} \rightarrow e^{+}e^{-} b\overline{b} \\
& \text{ Signal 2:} \\
& \quad \quad W fusion && e^{+}e^{-} \rightarrow \overline{\nu_{e}} \nu_{e} b\overline{b} 
\end{align*}
Here, $l$ and $\overline{l}$ are $e^-,\mu^-$ and $e^+,\mu^+$, respectively. The corresponding background processes analysed here are as follows:
\begin{align*}
& \text{ For signal 1:}\\
i)&   \quad \quad  e^{+}e^{-} \rightarrow ZZ \rightarrow l \overline{l} J J,\\
ii)& \quad \quad  e^{+}e^{-} \rightarrow e^{+}e^{-} Z  \rightarrow e^{+}e^{-} J J, \\
iii)& \quad \quad  e^{+}e^{-} \rightarrow t \overline{t} \rightarrow W^{+} J W^{-} J \rightarrow l \overline{l} J J \nu_{l} \overline{\nu_{l}}, \\
& \text{ For signal 2:}\\
& \quad \quad  e^{+}e^{-} \rightarrow J J, 
\end{align*}
here, $J$ represents the quark and antiquark: $J = d, \overline{d}, u, \overline{u}, s, \overline{s}, c, \overline{c}, b, \overline{b}$. The transverse momentum ($P_T$), pseudo rapidity ($\eta$) and invariant mass ($M_{inv}$) distributions of the final state particles were investigated by using CalcHEP program in order to find the cut values to distinguish the signal from the background in the FCC-ee collider with a center of mass energy of 240 GeV. The background $iii$ process corresponding to Signal 1 is not included in the calculations for 240 GeV, as it starts to contribute at 350 GeV and greater center of mass energies. Because the transverse momentum, pseudo rapidity, and invariant mass distributions of the final state particles in the signal and background processes will exhibit similar behavior for other colliders, the cut values obtained can be used for CEPC, ILC, and CLIC.
\begin{figure}[!t]
\centering
\includegraphics[width=0.48\textwidth]{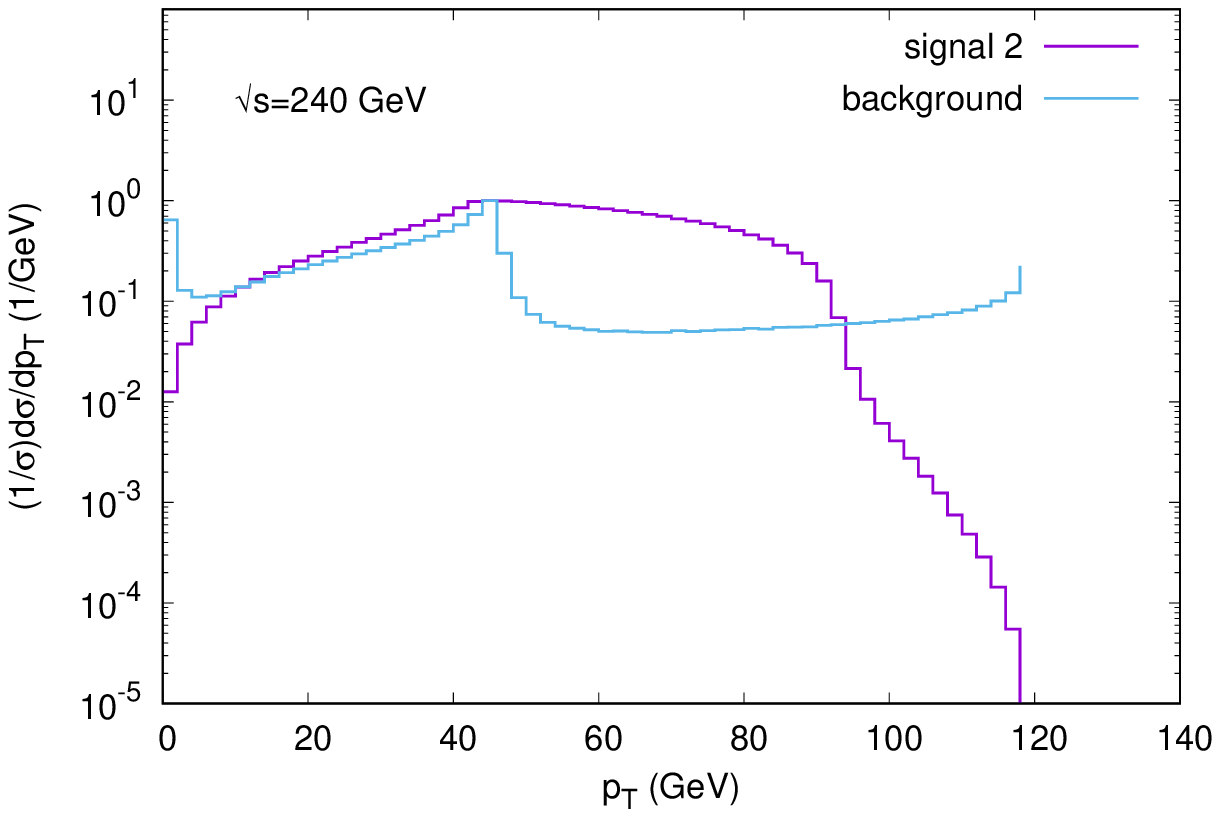}
\caption{Transverse momentum distribution plots for the $b/\overline{b}$ and $J/J$ final state particles of signal 2 and the corresponding bacground processes in FCC-ee collider with 240 GeV center of mass energy. }
\label{figure5}
\end{figure}
Transverse momentum distribution plots for the final state particles of signal 1 and the corresponding background processes \textit{i} and \textit{ii} are shown in  Figure \ref{figure4}, while the graphs of signal 2 are shown in Figure \ref{figure5}. As can be seen from Figure \ref{figure4} and \ref{figure5}, when a transverse momentum cut of 35 GeV is applied to the $e^-$, $e^+$, $\mu^-$, $\mu^+$, and two jets ($J$)  in the final state particles of signal 1 and signal 2  and the corresponding background processes, the signal will almost not change, but the background will be significantly reduced.

\begin{figure}[!b]
\centering
\begin{minipage}{0.48\textwidth}
\includegraphics[width=\textwidth]{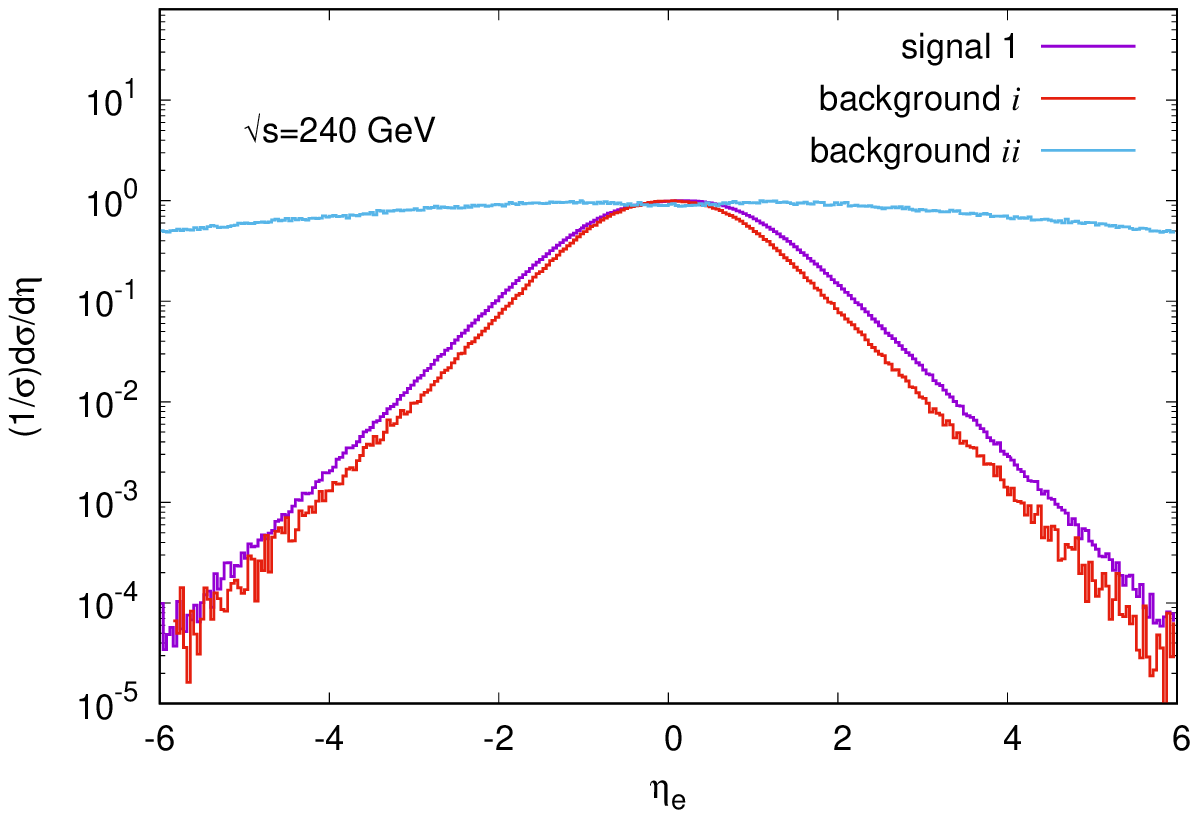}
\end{minipage}\hfill
\begin{minipage}{0.48\textwidth}
\includegraphics[width=\textwidth]{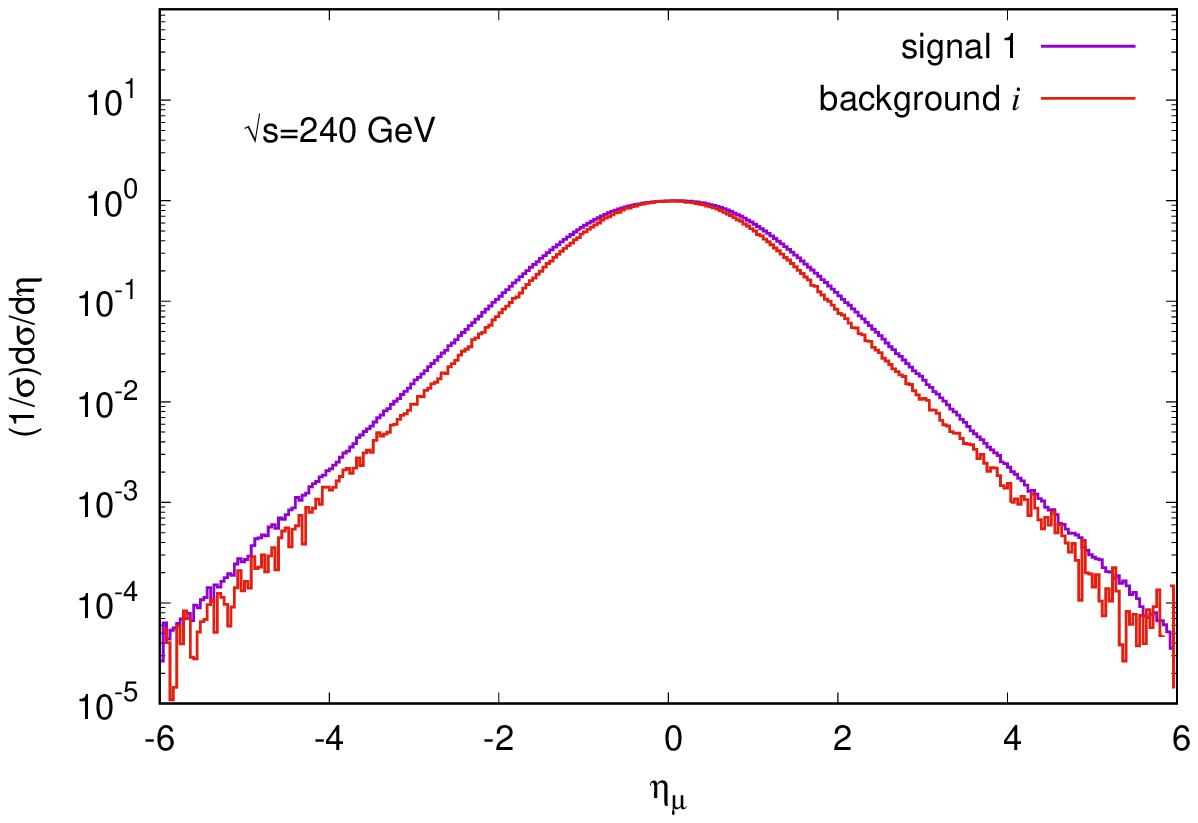}
\end{minipage}\par
\vskip\floatsep
\includegraphics[width=0.48\textwidth]{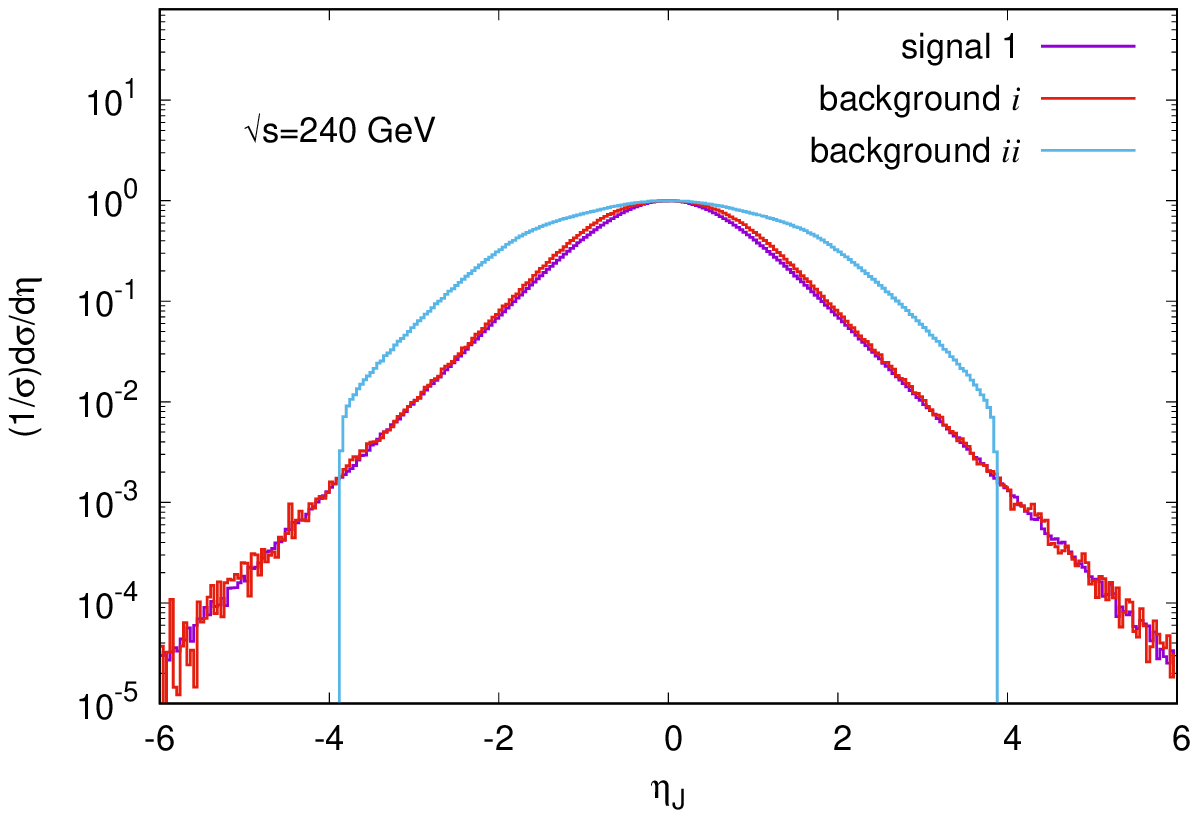}
\caption{Pseudorapidity distribution plots for the $e^-/e^+$ (upper left ), $\mu^-/\mu^+$ (upper right) and $J/J$ (bottom) final state particles of signal 1 and the corresponding bacground processes in FCC-ee collider with 240 GeV center of mass energy. }
\label{figure6}
\end{figure}

\begin{figure}[!t]
\centering
\includegraphics[width=0.48\textwidth]{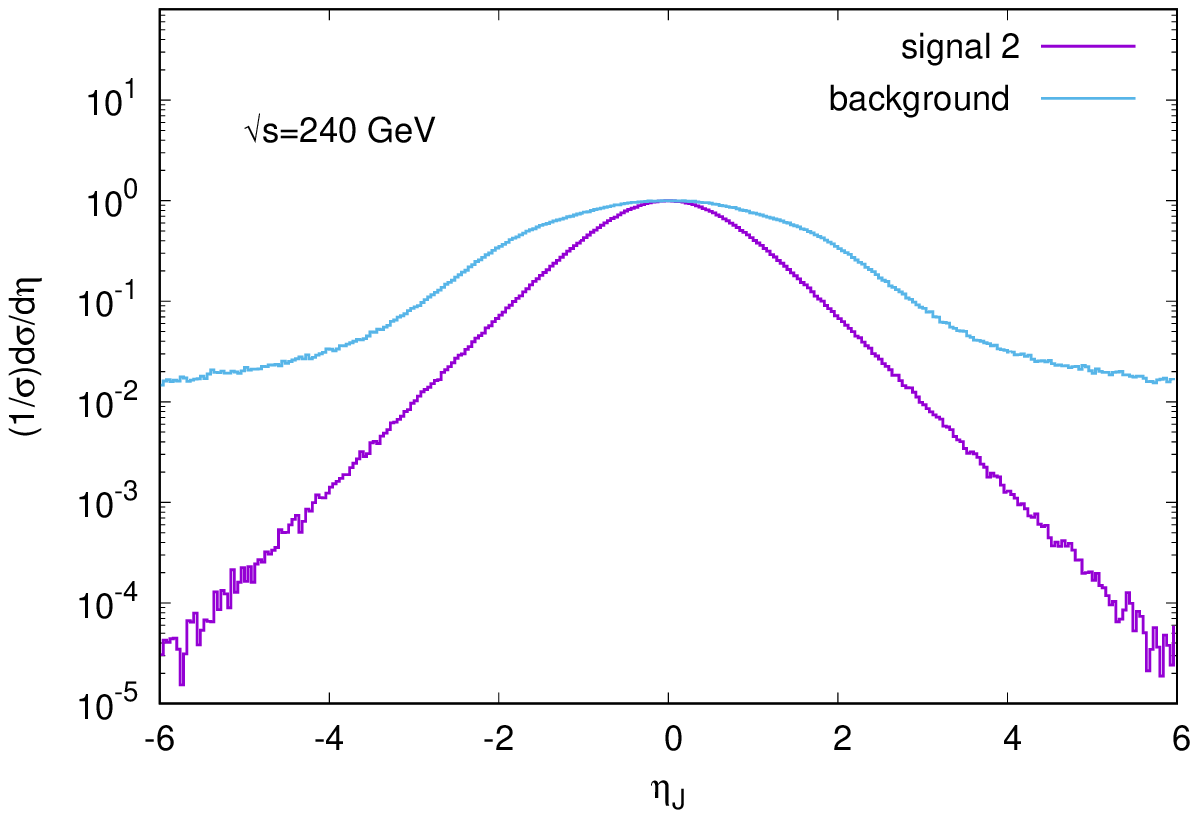}
\caption{Pseudorapidity distribution plots for the $b/\overline{b}$ and $J/J$ final state particles of signal 2 and the corresponding bacground processes in FCC-ee collider with 240 GeV center of mass energy. }
\label{figure7}
\end{figure}

Pseudorapidity plots for signal 1, signal 2, and the corresponding backgrounds are shown in Figure \ref{figure6} and \ref{figure7}. As can be seen from the figures, cut regions of $-2.5 < \eta_{J,J} < 2.5$, $-2.5 < \eta_{e^-,e^+} < 2.5$, $-2.5 < \eta_{\mu^-,\mu^+} < 2.5$  will be appropriate for  $e^-$, $e^+$, $\mu^-$, $\mu^+$ and two jets (J)  in the final state particles of signal 1 and signal 2  and the corresponding background processes. 

An $E_T^{miss}$ cut value  of $>$15 GeV was also used for neutrinos in our calculations.

\begin{figure}[h]
\centering
\begin{minipage}{0.48\textwidth}
\includegraphics[width=\textwidth]{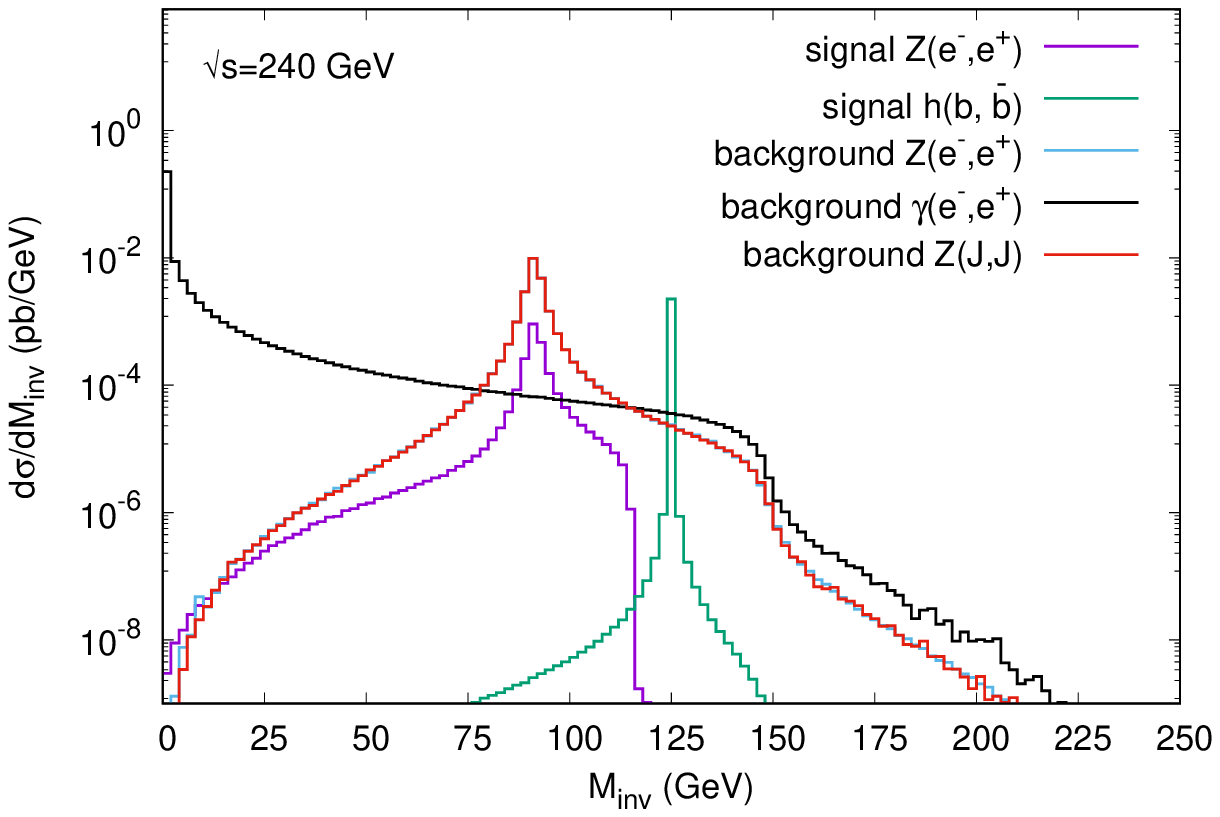}
\end{minipage}\hfill
\begin{minipage}{0.48\textwidth}
\includegraphics[width=\textwidth]{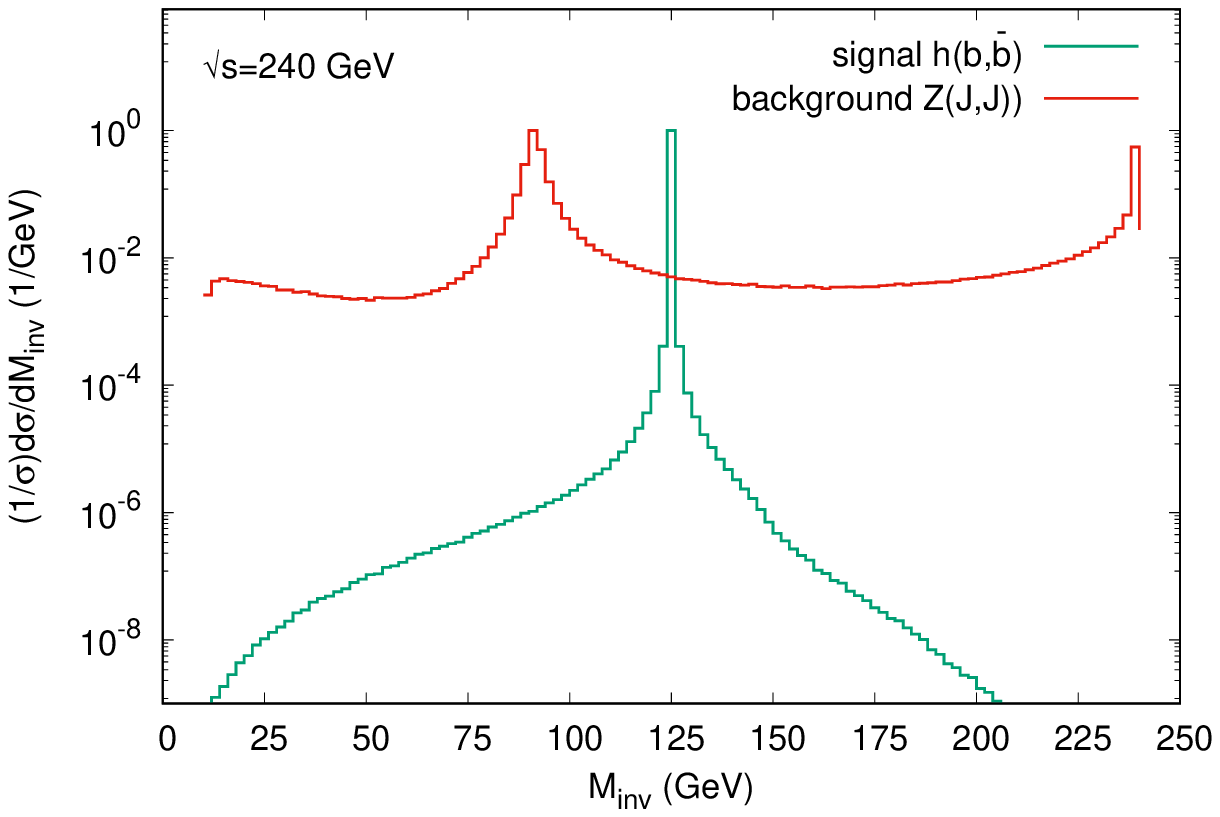}
\end{minipage}
\caption{Invariant mass plots for the signal 1 (left) and signal 2 (right) and the corresponding bacground processes in FCC-ee collider with 240 GeV center of mass energy.}
\label{figure8}
\end{figure}

Invariant mass distribution plots for signal 1, signal 2 and their corresponding background processes are shown in Figure \ref{figure8}. As can be seen from the figures, in the calculations, it would be appropriate to exclude the 80 GeV $< M_{inv}(e^-,e^+) <$100 GeV and 80 GeV $< M_{inv}(\mu^-,\mu^{+}) <$ 100 GeV regions for the $l \overline{l}$ final states in signal and background processes. In addition, only the 115 GeV $< M_{inv}(J, J) <$135 GeV region was included in the calculations for two final jet states in the signal and background processes. These included and excluded invariant mass regions allow the signal to be distinguished from the background.

In addition to these cut values, the separation cuts of $\Delta R (l,J)>$0.5 and $\Delta R (\overline{l},J)>$0.5 distinguish the final state leptons and antileptons from the jets, while the $\Delta R(J,J) >$0.5 separation cut was used to distinguish the final state jets from each other.

All the cut values obtained are listed in Table~\ref{table2}, and these cut values were used in the calculations for the four colliders. In addition to the cut values in Table ~\ref{table2}, because the Higgs boson decays to $b\overline{b}$ in our signal processes, it is possible to further reduce the background cross section value using the b-tagging method \cite{citation27}: 68\% is used for the b-tagging identification rate, and a 1\% ratio is used for misidentification rate with light quarks as b quarks.
\begin{table}[!b]
\caption{Cut values}
\medskip
\centering\renewcommand{\arraystretch}{1.2}
\begin{tabular}{|c|}
\hline
$E^{miss}_T(\nu_{l}, \overline{\nu_{l}})> $ 15 GeV            \\ \hline
$P_T(l,\overline{l})>$ 35 GeV                                   \\ \hline
$P_T(J)>$ 35 GeV                                         \\ \hline
-2.5 $<\eta (l,\overline{l})<$ 2.5                           \\ \hline
-2.5 $<\eta (J)<$ 2.5                                 \\ \hline
80 GeV $<M_{inv}(l,\overline{l})<$ 100 GeV region is excluded \\ \hline
115 GeV$<M_{inv}(J,J)<$ 135 GeV region is included   \\ \hline
$\Delta R (l,J)>$ 0.5                                           \\ \hline
$\Delta R (\overline{l},J)>$ 0.5                                      \\ \hline
$\Delta R (J,J)>$ 0.5                                           \\ \hline
\end{tabular}
\label{table2}
\end{table}
The following equation is used to calculate the significance of the obtained data: 
\begin{equation}
\label{equation1} 
\mathcal{S}=\sqrt{2((s+b)\ln(1+s/b)-s)}
\end{equation}
where $s$ and $b$ represent signal and background events, respectively \cite{citation32}. Cross-sections, event rates, and significance values were calculated for the signal and background processes using the cut values in Table~\ref{table2}, b-tagging method, and nominal integrated luminosity given in Table~\ref{table1}. The event rates and significance values of the signals and corresponding backgrounds are obtained for four future lepton colliders. The numerical results are given in Table~\ref{table3}-\ref{table6}. The abbreviations used in the tables are: Sg CS (signal cross-section), Bg CS ( background cross-section),$ \mathcal{L}$ (integrated luminosity), No. SgE (number of signal events) and No. BgE (number of background events).

\begin{table}[!t]
\caption{Cross sections, number of events and the significance values for CEPC.}
\medskip
\centering\renewcommand{\arraystretch}{1.2}
\begin{tabular}{|c|c|l|l|l|l|l|l|}
\hline
Colliders   & Processes  & \multicolumn{1}{c|}{\begin{tabular}[c]{@{}c@{}}Sg CS\\  ($pb$)\end{tabular}} & \multicolumn{1}{c|}{\begin{tabular}[c]{@{}c@{}}Bg CS\\   ($pb$)\end{tabular}} & \multicolumn{1}{c|}{\begin{tabular}[c]{@{}c@{}}$\mathcal{L}$\\ ($pb^{-1}$)\end{tabular}} & No. SgE & No.BgE & $\mathcal{S}$ \\ \hline
\multirow{4}{*}{\begin{tabular}[c]{@{}c@{}}CEPC\\ (240 GeV)\end{tabular}} 
& Signal 1  & 9.91$\times 10^{-5}$ & 2.38$\times 10^{-4}$   & \multirow{4}{*}{6$\times 10^{5}$}    & 59.5  & 142.8      & 4.7 \\ \cline{2-4} \cline{6-8} 
& \begin{tabular}[c]{@{}c@{}}Signal 1\\ (with b-tagging)\end{tabular} & 6.72$\times 10^{-5}$   & 3.68$\times 10^{-5}$                                                                                                & &40.32    & 22.08        &7          \\ \cline{2-4} \cline{6-8} 
& Signal 2     & 1.24$\times 10^{-2}$                                                                                                     & 5.22$\times 10^{-1}$                                                                                                                                                                                                            &      & 7440        & 313200       & 13.24  \\ \cline{2-4} \cline{6-8} 
 & \begin{tabular}[c]{@{}c@{}}Signal 2\\ (with b-tagging)\end{tabular} &  8.45$\times 10^{-3}$                                                                                                                                                                                                          & 7.17$\times 10^{-2}$                                                                                                                                                                                                            &      & 5070       & 43020       & 23.98  \\ \hline
\end{tabular}
\label{table3}
\end{table}
\begin{table}[!b]
\caption{Cross sections, number of events and the significance values for FCC-ee.}
\medskip
\centering\renewcommand{\arraystretch}{1.2}
\begin{tabular}{|c|c|l|l|l|l|l|l|}
\hline
Colliders   & Processes  & \multicolumn{1}{c|}{\begin{tabular}[c]{@{}c@{}}Sg CS\\  ($pb$)\end{tabular}} & \multicolumn{1}{c|}{\begin{tabular}[c]{@{}c@{}}Bg CS\\   ($pb$)\end{tabular}} & \multicolumn{1}{c|}{\begin{tabular}[c]{@{}c@{}}$\mathcal{L}$\\ ($pb^{-1}$)\end{tabular}} & No. SgE & No.BgE & $\mathcal{S}$ \\ \hline
\multirow{4}{*}{\begin{tabular}[c]{@{}c@{}}FCC-ee\\ (240 GeV)\end{tabular}} 
& Signal 1  & 9.98$\times 10^{-5}$ & 2.36$\times 10^{-4}$   & \multirow{4}{*}{1.7$\times 10^{6}$}    & 169.7  & 401.2      & 7.95 \\ \cline{2-4} \cline{6-8} 
& \begin{tabular}[c]{@{}c@{}}Signal 1\\ (with b-tagging)\end{tabular} & 6.79$\times 10^{-5}$   & 3.67$\times 10^{-5}$                                                                                                & &115.4    & 62.4       & 11.9          \\ \cline{2-4} \cline{6-8} 
& Signal 2     & 1.25$\times 10^{-2}$                                                                                                     & 5.37$\times 10^{-1}$                                                                                                                                                                                                            &      & 21250       & 912900       & 22.15  \\ \cline{2-4} \cline{6-8} 
 & \begin{tabular}[c]{@{}c@{}}Signal 2\\ (with b-tagging)\end{tabular} &  8.53$\times 10^{-3}$                                                                                                                                                                                                          & 7.38$\times 10^{-2}$                                                                                                                                                                                                            &      & 14501       & 125460       & 40.18  \\ \hline 
\end{tabular}
\label{table4}
\end{table}
\begin{table}[h!]
\caption{Cross sections, number of events and the significance values for ILC.}
\medskip
\centering\renewcommand{\arraystretch}{1.2}
\begin{tabular}{|c|c|l|l|l|l|l|l|}
\hline
Colliders   & Processes  & \multicolumn{1}{c|}{\begin{tabular}[c]{@{}c@{}}Sg CS\\  ($pb$)\end{tabular}} & \multicolumn{1}{c|}{\begin{tabular}[c]{@{}c@{}}Bg CS\\   ($pb$)\end{tabular}} & \multicolumn{1}{c|}{\begin{tabular}[c]{@{}c@{}}$\mathcal{L}$\\ ($pb^{-1}$)\end{tabular}} & No. SgE & No.BgE & $\mathcal{S}$ \\ \hline
\multirow{4}{*}{\begin{tabular}[c]{@{}c@{}}ILC\\ (250 GeV)\end{tabular}} 
& Signal 1  & 1.33$\times 10^{-4}$ & 2.9$\times 10^{-4}$   & \multirow{4}{*}{1.35$\times 10^{5}$}    & 17.95  & 39.15      & 2.68 \\ \cline{2-4} \cline{6-8} 
& \begin{tabular}[c]{@{}c@{}}Signal 1\\ (with b-tagging)\end{tabular} & 9.04$\times 10^{-5}$   & 4.34$\times 10^{-5}$     &  &12.2    & 5.86       & 4.03         \\ \cline{2-4} \cline{6-8} 
& Signal 2     & 1.3$\times 10^{-2}$                                                                                                     & 5.33$\times 10^{-1}$                                                                                                                                                                                                            &      & 1755      & 71955       & 6.51  \\ \cline{2-4} \cline{6-8} 
 & \begin{tabular}[c]{@{}c@{}}Signal 2\\ (with b-tagging)\end{tabular} &  8.82$\times 10^{-3}$                                                                                                                                                                                                          & 7.32$\times 10^{-2}$                                                                                                                                                                                                            &      & 1190      & 9882      & 11.74  \\ \hline 
 
\multirow{4}{*}{\begin{tabular}[c]{@{}c@{}}ILC\\ (500 GeV)\end{tabular}} 
& Signal 1  & 1.41$\times 10^{-3}$ & 1.7$\times 10^{-3}$   & \multirow{4}{*}{1.8$\times 10^{5}$}    & 253.8  & 306      & 12.98 \\ \cline{2-4} \cline{6-8} 
& \begin{tabular}[c]{@{}c@{}}Signal 1\\ (with b-tagging)\end{tabular} & 9.57$\times 10^{-4}$   & 3.32$\times 10^{-4}$     &  &172.3    & 59.8       & 16.88        \\ \cline{2-4} \cline{6-8} 
& Signal 2     & 2.86$\times 10^{-2}$                                                                                                     & 1.13$\times 10^{-1}$                                                                                                                                                                                                            &      & 5148     & 20340      & 34.71  \\ \cline{2-4} \cline{6-8} 
 & \begin{tabular}[c]{@{}c@{}}Signal 2\\ (with b-tagging)\end{tabular} &  1.94$\times 10^{-2}$                                                                                                                                                                                                          & 1.56$\times 10^{-2}$                                                                                                                                                                                                            &      & 3492     & 2808      & 56.54  \\ \hline 
\end{tabular}
\label{table5}
\end{table}
\begin{table}[!h]
\caption{Cross sections, number of events and the significance values for CLIC.}
\medskip
\centering\renewcommand{\arraystretch}{1.2}
\begin{tabular}{|c|c|l|l|l|l|l|l|}
\hline
Colliders   & Processes  & \multicolumn{1}{c|}{\begin{tabular}[c]{@{}c@{}}Sg CS\\  ($pb$)\end{tabular}} & \multicolumn{1}{c|}{\begin{tabular}[c]{@{}c@{}}Bg CS\\   ($pb$)\end{tabular}} & \multicolumn{1}{c|}{\begin{tabular}[c]{@{}c@{}}$\mathcal{L}$\\ ($pb^{-1}$)\end{tabular}} & No. SgE & No.BgE & $\mathcal{S}$ \\ \hline
\multirow{4}{*}{\begin{tabular}[c]{@{}c@{}}CLIC\\ (380 GeV)\end{tabular}} 
& Signal 1  & 6.44$\times 10^{-4}$ & 1.08$\times 10^{-3}$   & \multirow{4}{*}{1.5$\times 10^{5}$}    & 96.6  & 162      & 6.98 \\ \cline{2-4} \cline{6-8} 
& \begin{tabular}[c]{@{}c@{}}Signal 1\\ (with b-tagging)\end{tabular} & 4.38$\times 10^{-4}$   & 2.53$\times 10^{-4}$     &  &65.7    & 37.95       & 8.76        \\ \cline{2-4} \cline{6-8} 
& Signal 2     & 1.7$\times 10^{-2}$                                                                                                     & 2.11$\times 10^{-1}$                                                                                                                                                                                                            &      & 2550    & 31650      &14.14  \\ \cline{2-4} \cline{6-8} 
 & \begin{tabular}[c]{@{}c@{}}Signal 2\\ (with b-tagging)\end{tabular} &  1.16$\times 10^{-2}$                                                                                                                                                                                                          & 2.9$\times 10^{-2}$                                                                                                                                                                                                            &      & 1740     & 4350      & 24.86  \\ \hline   
 
\multirow{4}{*}{\begin{tabular}[c]{@{}c@{}}CLIC\\ (1500 GeV)\end{tabular}} 
& Signal 1  & 1.95$\times 10^{-3}$ & 1.47$\times 10^{-3}$   & \multirow{4}{*}{3.7$\times 10^{5}$}    & 721.5 & 543.9      & 26.34 \\ \cline{2-4} \cline{6-8} 
& \begin{tabular}[c]{@{}c@{}}Signal 1\\ (with b-tagging)\end{tabular} & 1.32$\times 10^{-3}$   & 2.34$\times 10^{-4}$     &  &488.4    & 86.58       & 36.64       \\ \cline{2-4} \cline{6-8} 
& Signal 2     & 1.03$\times 10^{-1}$                                                                                                     & 9.22$\times 10^{-3}$                                                                                                                                                                                                            &      & 38110    & 3411      & 362  \\ \cline{2-4} \cline{6-8} 
 & \begin{tabular}[c]{@{}c@{}}Signal 2\\ (with b-tagging)\end{tabular} &  6.99$\times 10^{-2}$                                                                                                                                                                                                          & 1.27$\times 10^{-3}$                                                                                                                                                                                                            &      & 25863     & 470      & 400  \\ \hline    
 
\multirow{4}{*}{\begin{tabular}[c]{@{}c@{}}CLIC\\ (3000 GeV)\end{tabular}} 
& Signal 1  & 6.01$\times 10^{-4}$ & 8.47$\times 10^{-4}$   & \multirow{4}{*}{5.9$\times 10^{5}$}    & 355 & 499.7     & 14.38 \\ \cline{2-4} \cline{6-8} 
& \begin{tabular}[c]{@{}c@{}}Signal 1\\ (with b-tagging)\end{tabular} & 4.09$\times 10^{-4}$   & 1.32$\times 10^{-4}$     &  &241    & 77.8      & 20.44       \\ \cline{2-4} \cline{6-8} 
& Signal 2     & 1.61$\times 10^{-1}$                                                                                                     &2.72$\times 10^{-3}$                                                                                                                                                                                                            &      & 94990    & 1605      & 776  \\ \cline{2-4} \cline{6-8} 
 & \begin{tabular}[c]{@{}c@{}}Signal 2\\ (with b-tagging)\end{tabular} &  1.09$\times 10^{-1}$                                                                                                                                                                                                          & 3.74$\times 10^{-4}$                                                                                                                                                                                                            &      & 64310     & 221      & 777  \\ \hline     
\end{tabular}
\label{table6}
\end{table}

\section{Conclusion}
After the discovery of the Higgs particle, precise measurements of the Higgs properties became an important step forward for future research in particle physics. Electron positron colliders to be installed for this purpose have unique capabilities for the measurement of Higgs boson parameters, including the Higgs total cross section of production processes, decay width, branching rates and determination of Higgs couplings. In this study, the Higgsstrahlung and W and Z fusion processes were examined, and the data obtained are presented in graphs and tables for four different electron-positron colliders. The production cross-sections for each process and additionally cross-sections for various final state backgrounds were calculated. In the calculations, we attempted to reduce the background by transverse momentum, pseudo rapidity, invariant mass, cone-angle constraints, and the b-tagging method. Significance calculations were performed by determining the number of events related to the production processes and the background for each collider. The values are listed in Table~\ref{table3}-\ref{table6}.

When the results are examined in Table~\ref{table5} , it is seen that the desired significance value for Signal 1 cannot be reached at the luminosity value given for ILC – 250 GeV. For Signal 1 processes to be observed in the ILC-250 GeV, the collider needs to accumulate data for a longer period of time. Again, at the end of one year, it was seen that the statistical significance value of $5\sigma$ would be reached after the b-tagging method for the Signal 1 processes in the CEPC collider. Therefore, the CEPC collider will enable the properties of the Higgs boson to be investigated precisely through Signal 1 processes. It is seen that at the end of 1 year in the FCC-ee collider, a significance value of 7.95 will be reached without b-tagging and a high significance value of 11.9 can be reached by using b-tagging. This shows that FCC-ee will be more advantageous than ILC-250 GeV and CEPC 250 GeV colliders for investigating Higgs boson properties through the Signal 1 group around these center of mass energies (240-250 GeV).
In the ILC-500 GeV and CLIC-380-1500-3000 GeV colliders, results well above the desired significance value can be obtained for signal 1 processes, even without the b-tagging. Therefore, the properties of the Higgs boson through Signal 1 processes can be studied with precision in colliders other than the ILC-250 GeV collider. Since the results obtained for the Signal 2 process are greater than 5 significance values, the properties of the Higgs boson can be studied precisely for all colliders through this channel.

As a result, in future lepton colliders, the Higgs boson can be observed with high event rates via Higgsstrahlung, W and Z fusion. Thus, electron–positron colliders can precisely measure the properties of the Higgs boson.
\section*{Acknowledgment}
We would like to thank Professor Dr Inanc Sahin for his suggestions.

\end{document}